%% file: depowl.tex
\documentclass[10pt,conference]{IEEEtran}
\IEEEoverridecommandlockouts
\usepackage{fancyhdr}
\usepackage{cite}
\usepackage{amsmath,amssymb,amsfonts}
\usepackage{graphicx}
\usepackage{wasysym}
\usepackage{xcolor}
\usepackage[normalem]{ulem}
\usepackage{enumitem}
\usepackage{multirow}
\usepackage{subcaption}
\usepackage{algorithm}
\usepackage{algpseudocode}
\usepackage[T1]{fontenc}
\usepackage{fontawesome}
\newcommand{\tabincell}[2]{\begin{tabular}{@{}#1@{}}#2\end{tabular}}
\newcommand{\commentty}[1]{{\color{red} \sf (TY: #1)}}

\newcommand{\Name}{\textit{DepOwl}}
\newcommand{\ignore}[1]{}
\def\BibTeX{{\rm B\kern-.05em{\sc i\kern-.025em b}\kern-.08em
    T\kern-.1667em\lower.7ex\hbox{E}\kern-.125emX}}
\begin{document}

\title{\Name{}: Detecting Dependency Bugs to Prevent Compatibility Failures
\thanks{We thank the anonymous reviewers for their insightful comments. 
We also thank Xin Peng, Bihuan Chen and Kaifeng Huang for their suggestions. 
This work was supported in part by 
NSFC No. 61872373;
National Key R\&D Program of China No. 2018YFB0204301; 
NSFC No. 61872375, U19A2060, 61802416; 
NSF grant CCF-1909085; 
and China Scholarship Council.}
}

\author{
\IEEEauthorblockN{
Zhouyang Jia\IEEEauthorrefmark{1}\IEEEauthorrefmark{2}, 
Shanshan Li\IEEEauthorrefmark{1}\IEEEauthorrefmark{3}\thanks{\IEEEauthorrefmark{3}Shanshan Li and Xiaodong Liu are the corresponding authors.}, 
Tingting Yu\IEEEauthorrefmark{2}, 
Chen Zeng\IEEEauthorrefmark{1},
Erci Xu\IEEEauthorrefmark{1},
Xiaodong Liu\IEEEauthorrefmark{1}\IEEEauthorrefmark{3},
Ji Wang\IEEEauthorrefmark{1},
Xiangke Liao\IEEEauthorrefmark{1}} 
\and
\IEEEauthorblockA{
\IEEEauthorrefmark{1}\textit{College of Computer Science} \\
\textit{National University of Defense Technology}\\
Changsha, China \\
\{jiazhouyang, shanshanli, zengchen15, xuerci, liuxiaodong, \\wj, xkliao\}@nudt.edu.cn}
\and
\IEEEauthorblockA{
\IEEEauthorrefmark{2}\textit{Department of Computer Science} \\
\textit{University of Kentucky}\\
Lexington, USA \\
tyu@cs.uky.edu}
}

\maketitle

\pagestyle{fancy}

\begin{abstract}
Applications depend on libraries to avoid reinventing the wheel.
Libraries may have incompatible changes during evolving. 
As a result, applications will suffer from compatibility failures.
There has been much research on addressing detecting incompatible changes in libraries, 
or helping applications co-evolve with the libraries.
The existing solution helps the latest application version work well against the latest library version as an afterthought.
However, end users have already been suffering from the failures and have to wait for new versions.
In this paper, we propose \Name{}, a practical tool helping users prevent compatibility failures.
The key idea is to avoid using incompatible versions from the very beginning.
We evaluated \Name{} on 38 known compatibility failures from StackOverflow, 
and \Name{} can prevent 35 of them.
We also evaluated \Name{} using the software repository shipped with Ubuntu-19.10. 
\Name{} detected 77 unknown dependency bugs, which may lead to compatibility failures.
\end{abstract}


\input{1-introduction}

\input{2-motivation}
\input{3-design}
\input{4-evaluation}

\input{5-discussion}

\input{6-relatedwork}

\input{7-conclusion}



\bibliographystyle{IEEEtran}
\bibliography{depowl}

\end{document}

%% file: 1-introduction.tex

\section{Introduction}\label{sec:intro}
Applications reuse as much existing code as possible for cost savings.
Existing code is often in the form of libraries,
which keep evolving and may introduce incompatible changes (e.g., changing interface signatures).
Misuses of library versions containing incompatible changes may lead to failures in applications.
We refer to these failures as \textit{compatibility failures}, or \textit{CFailures}.

\begin{figure}[tb]
{\centering \resizebox*{0.95\columnwidth}{!}
{\includegraphics{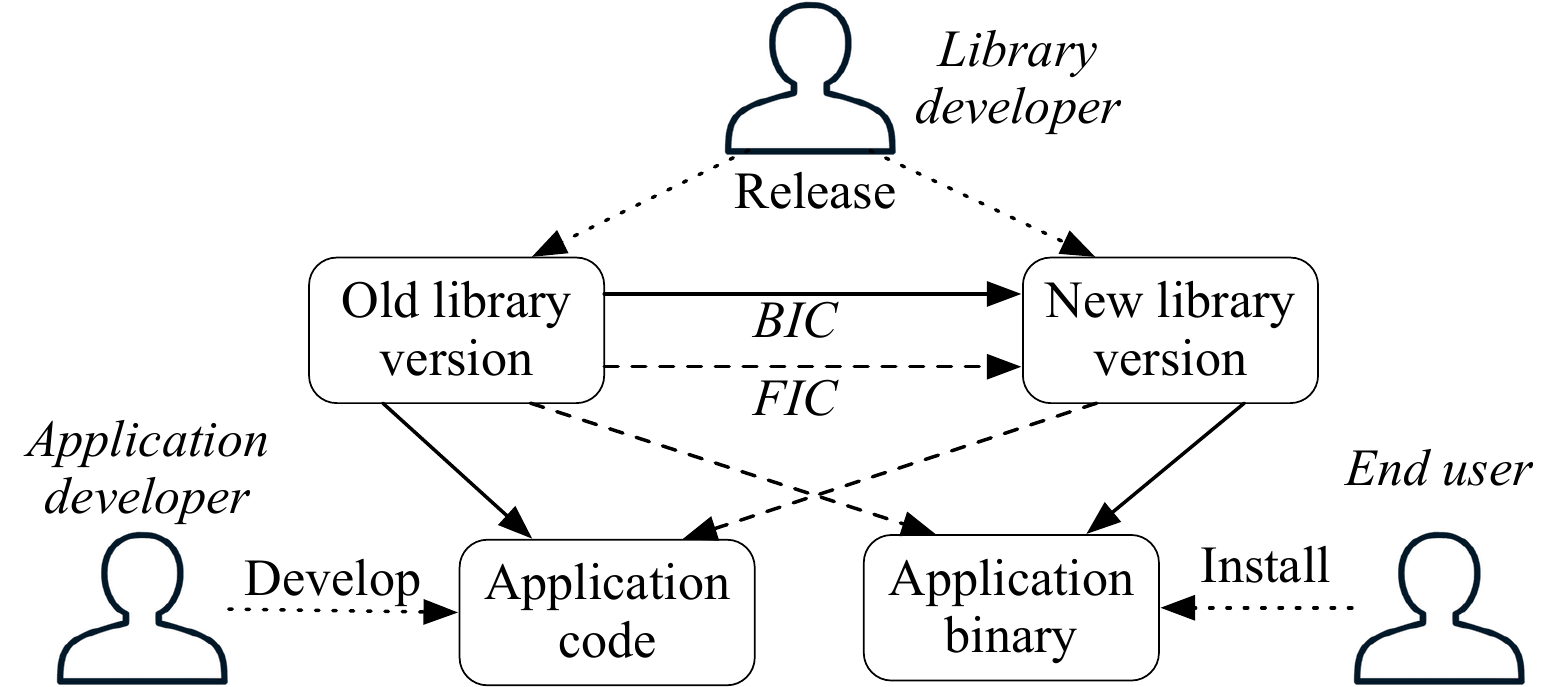}}
\caption{Incompatible changes cause \textit{CFailures}.  \small{\emph{The solid and dashed lines show how \textit{BIC} (backward incompatible changes) and \textit{FIC} (forward incompatible changes) cause \textit{CFailures}, respectively.}}}
\label{fig:failures}}
\vspace*{-12pt}
\end{figure}

A \textit{CFailure} involves three roles: library developers, application developers, and end users 
(\textit{library} and \textit{application} are relative concepts as an application itself may be a library for anther application).
As shown in Figure~\ref{fig:failures},
library developers release two versions  containing incompatible changes.
The changes are classified into two types:
\textit{backward incompatible change (BIC)} (e.g., removing an interface),
and \textit{forward incompatible change (FIC)} (e.g., adding an interface).
The solid (dashed) lines show how a \textit{BIC} (an \textit{FIC}) causes \textit{CFailures}:
if application developers develop an application based on the old (new) library version,
end users may suffer from \textit{CFailures} when linking the application to the new (old) library version.
In either case, the incompatible change causes \textit{CFailures}. 

\ignore{
The incompatible changes in a library can introduce
\emph{dependency bugs}. A dependency bug happens when the version range (usable
by the application) specified in the library contains incompatible
versions and can cause compatibility failures when interacting with the application. 
\commentty{Please double-check if I made the correct argument.}
}

When incompatible changes happened, the three roles can prevent \textit{CFailures} with different solutions:
1)~library developers can undo the changes in the latest version;
2)~application developers can update the application to adapt the changes;
3)~end users can avoid using the incompatible library versions. 
There has been some research on detecting library changes
~\cite{ 8330249, Ponomarenko2012, 6062100,
10.5555/2337223.2337265, mezzetti_et_al:LIPIcs:2018:9212,
10.1145/3236024.3275535}. These techniques focus on suggesting
incompatible changes for library developers (i.e., the first solution).
There has also been some work on detecting incompatible API usages in applications~\cite{
10.1145/3238147.3238185, 10.1145/3213846.3213857, 8812128, 6619503},
or helping applications adapt library changes~\cite{
10.1145/1094811.1094832, 1553570, 10.1145/1108792.1108818, 4359473}.
These techniques focus on helping application developers 
update the application (i.e., the second solution).
In either of the above solutions, 
end users may have already suffered from \textit{CFailures} and have to wait for new library/application versions.
The third solution, on the other hand, is more light-weighted --- 
end users can avoid \textit{CFailures} from the very beginning without having to see the 
\textit{CFailures} occur.  However, 
there exists no research that can achieve this goal by helping users automatically select  compatible library versions. 

Some industrial settings use dependency management systems (\textit{DMSs})
that can help users select right library versions.
Examples include \textit{dnf}~\cite{Fedora2019Using} in RPM-based Linux distributions and 
\textit{apt}~\cite{Ubuntu2019Apt} in Debian-based Linux distributions.
However, \textit{DMSs} have several practical limitations (more details in Section~\ref{sec:limitation}):

\begin{enumerate}
\item \textit{DMSs} require manual inputs from either application or library developers, 
which can be tedious and error-prone.
For example, \textit{dnf} requires application developers to specify version ranges of required libraries.
\textit{apt} asks library developers to maintain a symbol list provided by the library. 

\item Manual inputs provided by developers may be outdated as the libraries evolve.
For example, application developers specified the version
range \textit{libfoo>=1.0}, 
after which \textit{libfoo-2.0} is released and backward incompatible to \textit{libfoo-1.0}.
The version range should have been updated to \textit{2.0>libfoo>=1.0}.

\item Developers may not comply with the requirements of the \textit{DMSs}.
For example, \textit{apt} requires libraries not to break backward compatibility in a package,
but library developers may unintentionally introduce incompatibilities
since there is no mechanism to guarantee the requirement.

\end{enumerate}

Since \textit{DMSs} depend on version ranges specified in specification files 
(e.g., the \textit{control} file used by \textit{apt}, or the \textit{.spec} file used by \textit{dnf}) to resolve dependencies,
the above limitations may introduce incompatible versions being included in the version ranges.
In this case, we say there are \textit{dependency bugs} (or \textit{DepBugs}) in the specification files.

To address the limitations within \textit{DMSs}, 
we propose a new approach,  \Name{},  to detect \textit{DepBugs}
and prevent \textit{CFailures}.
\Name{} works at the binary level to check compatibility between
libraries and applications instead of
analyzing the API usage in source code of applications (e.g., \textit{compilers})\footnote{The current design of \Name{} focuses on C/C++ applications and libraries.}. 
This is advantageous 
for end users who prefer to install binary files without having to compile the source code.
For example, end users often use the command \textit{apt install} to download binary files.
The source-code level compatibility can not guarantee the compatibility
of the binary files installed by the users.

Specifically, given the binaries of a library and an application, \Name{} automatically checks if the application
is compatible to each version of the library, so it can help users
select the right library versions to prevent \textit{CFailures}. 
%
%
\Name{}  contains three major steps. In the first step, \Name{}  collects  
all potentially incompatible changes (e.g., add/remove/change interfaces) during the evolution of the library
(from an old version  to a new version),
including both \textit{BICs} and \textit{FICs}.  
Next, \Name{} checks if the API usage in the target application matches
the API definitions in either of the old and new library versions. 
If the change is a \textit{BIC} (\textit{FIC}) and the API usage matches the old (new) library version, 
the new (old) library version is regarded as an incompatible version.
In the third step, \Name{} compares the incompatible  version to all other library versions.
Any version that is both backward and forward \emph{compatible} to the incompatible 
version is also identified as an incompatible version.
Users can prevent \textit{CFailures} by avoiding using the
reported incompatible versions.

A common usage scenario of \Name{} is to serve as a plugin for \textit{DMSs}. 
Taking \textit{apt} as an example, in Debian-based Linux distributions, \textit{apt} helps users manage application dependencies. 
Each application contains a \textit{control} file indicating its required libraries and version ranges.
These ranges, however, may contain incompatible versions.
\Name{} is able to detect incompatible versions, 
so that \textit{apt} can avoid using incompatible versions when 
resolving dependencies, and users will be free of \textit{CFailures}.

We evaluated \Name{}'s ability in preventing both known and unknown \textit{CFailures}.  
We first evaluated \Name{} on 38 real-world known \textit{CFailures} from StackOverflow,
and \Name{} can prevent 35 of them.
We also applied  \Name{} to the software repository shipped with Ubuntu-19.10, the latest 
Ubuntu stable version at the time of writing. 
\Name{} detected 77 unknown \textit{DepBugs}, which may cause \textit{CFailures}.

In summary, the contributions of this paper are as follows:
\begin{enumerate}
\item We propose a lightweight solution to prevent \textit{CFailures} when incompatible changes happened in libraries.
Existing research work mainly focuses on fixing \textit{CFailures} in new versions, but can not prevent the \textit{CFailures}.
Industrial \textit{DMSs} can help users resolve dependencies, but still have limitations.

\item We design and implement \Name{}, a practical tool to detect \textit{DepBugs} and 
prevent \textit{CFailures}. \Name{} can collect incompatible changes 
in libraries, detect \textit{DepBugs} in applications, and suggest 
incompatible versions to help users prevent \textit{CFailures}.

\item \Name{} can prevent 35 out of 38 \textit{CFailures} selected from StackOverflow. 
and detect 77 \textit{DepBugs} in the repository shipped with Ubuntu-19.10.
\Name{} is more accurate compared with baseline methods, and requires no human efforts.
\end{enumerate}

%% file: 2-motivation.tex
\section{Existing \textit{DMSs} and Their Limitations}\label{sec:limitation}
Manual management of software dependencies is time-consuming and 
sometimes even error-prone, since an application may depend on many libraries, 
which keep evolving all the time.
In this regard, a common approach, especially in the open-source community, 
is to use a dependency management system (\textit{DMS}),
e.g., \textit{pip}~\cite{PyPA2019pip} for Python, 
\textit{Maven}~\cite{Apache2019Apache} for Java, 
\textit{npm}~\cite{npm2019npm} for JavaScript,
\textit{apt}~\cite{Ubuntu2019Apt} and \textit{dnf}~\cite{Fedora2019Using} in Linux distributions.

These \textit{DMSs} provide interfaces for developers to specify dependencies (i.e., the required
libraries and corresponding versions), as well as repositories that contain all libraries.
Developers manually specify dependencies, then the \textit{DMSs} can automatically download and 
install the libraries from the repositories.
For a required library, developers can specify a fixed version or a version range.
Using a fixed version is a reliable solution because it has little to virtually zero \textit{CFailures},
but it lacks flexibility because critical fixes in later versions of the library cannot be automatically 
included~\cite{10.1145/3133956.3134059}.
While using a version range increases flexibility since it can 
automatically include critical fixes in later versions of the library, 
but decreases its reliability because the later versions may also introduce \textit{CFailures}.
There is a tradeoff between flexibility and reliability in these two approaches.
Developers struggle to find the sweet spot~\cite{10.1109/MSR.2019.00061}.

\ignore{
For C/C++ projects, the libraries are usually in two different forms, i.e., static libraries
and dynamic (shared) libraries.
Static libraries put all the code into the executable file at compile time,
so programs will have faster execution times.
Programs with static libraries can be regarded as in an isolated environment,
where programs are built against a fixed library version, 
thus have little to virtually zero compatibility problem.
The builds are more deterministic, but critical fixes in later versions 
of the library cannot be automatically included~\cite{10.1145/3133956.3134059}.
On the other hand, dynamic libraries defer the linking process until a program starts running.
This feature significantly reduces the compiling time, dynamic linking load time, and the size 
of the executable program~\cite{Nickolas2018Shared}.
Programs with dynamic libraries can be regarded as in a shared environment,
where programs accept a version range of the target library, 
thus can automatically include critical fixes in later versions of the library.
But dynamic libraries may cause compatibility problems.
For example, program \textit{bar} relies on library \textit{libfoo-1}, 
while program \textit{cat} relies on \textit{libfoo-2}, which is not backward-compatible 
with \textit{libfoo-1}.
When users install \textit{cat} and updates \textit{libfoo-1} to \textit{libfoo-2}, 
\textit{bar} will suffer compatibility issues.
}

Most \textit{DMSs} leave this choice to application developers,
who can manually limit the version range of each required library.
Taking \textit{dnf} as an example, \textit{dnf} is the 
\textit{DMS} used in RPM-based Linux distributions like Fedora.
\textit{dnf} requires application developers to specify the 
required libraries and version ranges (e.g., \textit{ocaml>=3.08}),
which may be outdated: 
1)~The version ranges may be too large as libraries evolve.
For example, developers specify \textit{libfoo>=1.0} at first, 
after which \textit{libfoo-2.0} is released and backward incompatible with \textit{libfoo-1.0}.
In this case, the version range should be updated to \textit{2.0>libfoo>=1.0}.
2)~The version ranges may be too small as libraries evolve.
For example, developers specify \textit{libfoo<=1.0} at first, 
after which \textit{libfoo-2.0} is released and backward compatible with \textit{libfoo-1.0}.
In this case, the version range should be updated to \textit{libfoo<=2.0}.

To avoid these limitations, another solution is to maintain a \textit{symbols} file
by library developers.
This solution is applied in \textit{apt}, the \textit{DMS} in Debian-based Linux distributions like Ubuntu.
According to Debian policy~\cite{Debian2019Shared}: 
1)~"ABI (Application Binary Interface) changes that are not backward-compatible 
require changing the \textit{soname}~\cite{Wikipedia2019soname} of the library";
2)~"A shared library must be placed in a different package whenever its \textit{soname} changes".
It means that two library versions should be placed in two library packages, 
when the versions are backward incompatible.
These two packages, to some degree, can be regarded as two different libraries, 
e.g., \textit{libssl1.0.0} and \textit{libssl1.1}.
Library developers are required to maintain a 
\textit{symbols} file~\cite{Debian2019Shared}, in which
each line contains a symbol provided by the library, as well as the minimal version 
that the symbol is introduced.
Then, the version range of this library can be inferred automatically by extracting 
symbols used by an application. 
The minimal version of the version range is the maximum 
value of introducing versions of all used symbols.
The maximum version is not necessary since all versions are
backward compatible in one package.
Finally, the version range is used by \textit{apt} to help users 
manage dependencies.

\ignore{
\begin{figure}[tb]
{\centering \resizebox*{0.75\columnwidth}{!}
{\includegraphics{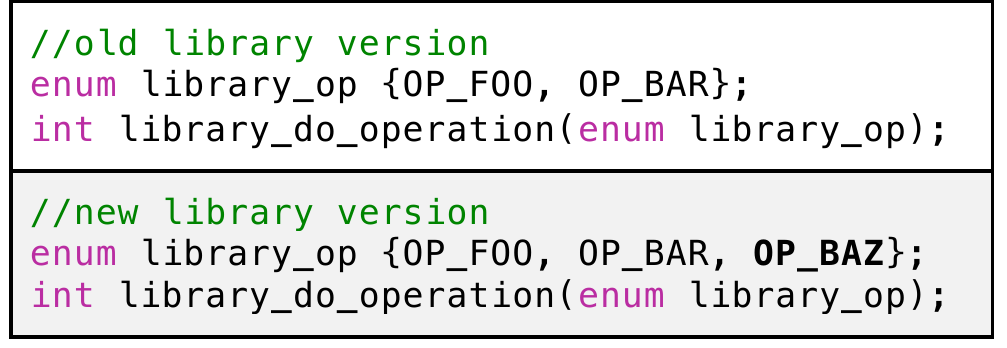}}
\caption{An example of library evolution~\cite{Debian2019Shared}}
\label{fig:library}}
\end{figure}
}

\begin{figure}[tb]
{\centering \resizebox*{0.95\columnwidth}{!}
{\includegraphics{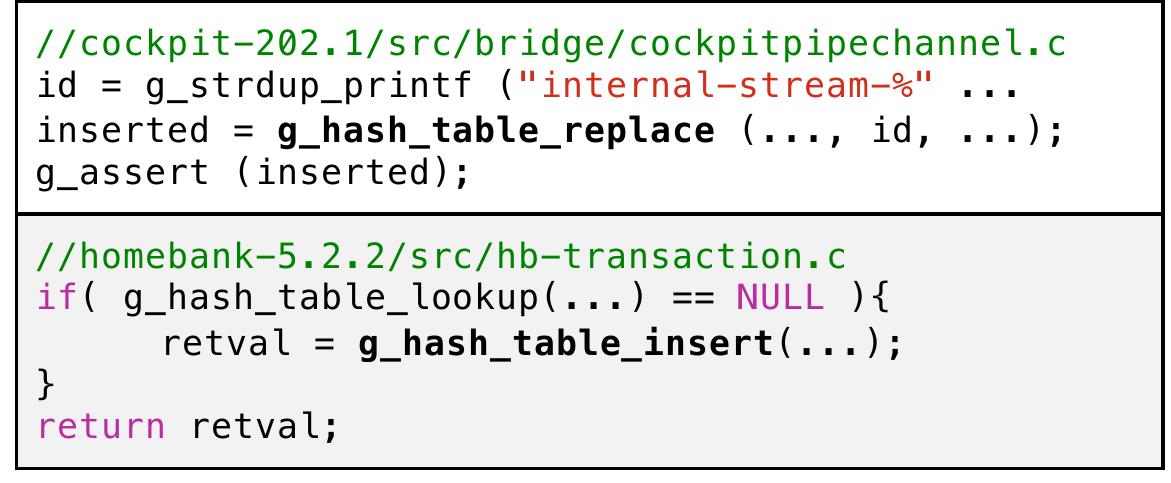}}
\caption{Example usages of library incompatible changes. \small{\emph{Both cockpit-202.1 and homebank-5.2.2 use return values of glib functions, which return void in some glib versions.}}}
\label{fig:example}}
\vspace*{-6pt}
\end{figure}

The above solution, however, is still limited since:
1)~There is no mechanism to guarantee that library developers comply with the policy. 
Library developers may unintentionally introduce ABI incompatibilities between two
versions, which have the same \textit{soname}.
Existing studies~\cite{10.1145/3236024.3275535, 6975655} show 26\%-33\% of library versions 
violate semantic versioning, meaning libraries frequently introduce 
incompatibilities during minor version changes.
%
2)~This solution only works for binary packages, since \textit{apt} needs to analyze 
binary files to extract symbols used by the application.
Application developers have to manually specify version ranges for source packages,
which do not have binary files.
In this case, \textit{apt} will suffer from the same limitations as \textit{dnf}.
3)~Library developers need to manually update the \textit{symbols} file when introducing forward incompatible changes.
For example, when a \textit{struct} type adds a \textit{field} in a new library version, 
the introducing version of all symbols using the \textit{struct}
must be increased to the version at which the new \textit{field} was introduced. 
Otherwise, a binary built against the new version of the library may be installed with 
a library version that does not support the new \textit{field}.
This is a common change during library evolutions, failing to update the introducing version
of any symbol will lead to \textit{DepBugs}.
We will show a real-world example in Section~\ref{sec:motivation}.


In summary, the \textit{DMSs} supporting version ranges may introduce \textit{DepBugs} ---
the ranges contain incompatible versions.
In this paper, we focus on detecting and fixing \textit{DepBugs} in the range-based \textit{DMSs}, 
so that applications can achieve higher reliability without affecting flexibility.

\begin{figure}[tb]
{\centering \resizebox*{1\columnwidth}{!}
{\includegraphics{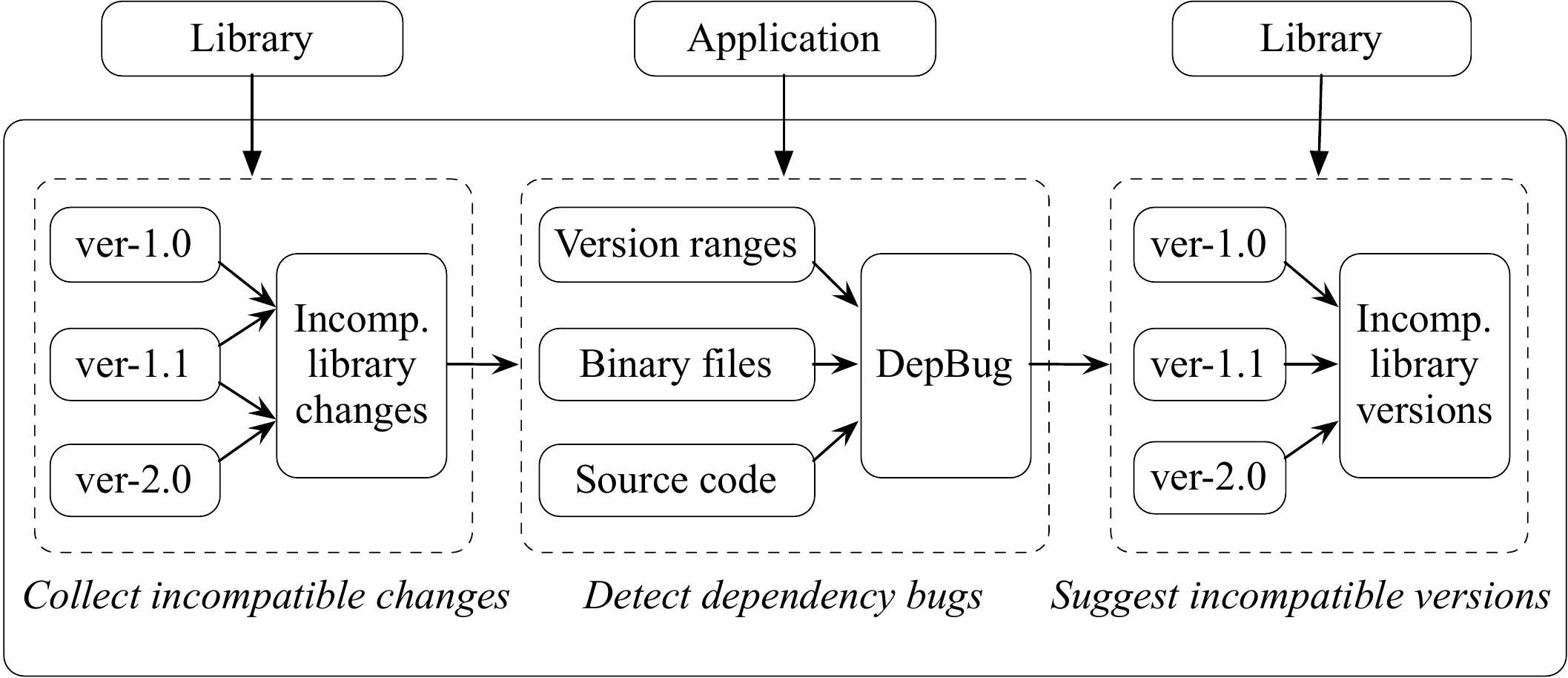}}
\caption{Overview of \Name{}. \small{\emph{\Name{} contains three major steps: 
collect incompatible changes, detect dependency bugs, and suggest incompatible versions.}}}
\label{fig:overview}}
\end{figure}

\section{Motivation and Overview of \Name{}}\label{sec:motivation}
In this section, we show a \textit{DepBug} example which motivates us to design \Name{}.
Based on the example, we introduce how \Name{} works at a high level.


\ignore{
The first example is \textit{zlib}, a library used for data compression.
The library developers remove the function \textit{gzgetc} 
(change \textit{gzgetc} to a macro for speed)
from \textit{zlib-1.2.5.1} to \textit{zlib-1.2.5.2}.
After that, the developers restore \textit{gzgetc} for compatibility
from \textit{zlib-1.2.5.2} to \textit{zlib-1.2.5.3}~\cite{Zlib2019ChangeLog}.
Any application using \textit{gzgetc} will suffer from 
\textit{CFailures} if depending on \textit{zlib-1.2.5.2}.
These changes affect many application packages in Ubuntu-19.10, e.g.,
\textit{ccache-3.7.3} (requires \textit{zlib>=1.1.4}) and
\textit{mathgl-2.4.4} (requires \textit{zlib>=1.2.0}).
Both the packages use the function \textit{gzgetc} and accept \textit{zlib-1.2.5.2},
which does not provide \textit{gzgetc}.
The root cause of this example is that the library developers break backward 
compatibility during minor version changes.
}

\textbf{Motivating example.} 
From \textit{glib-2.39.1} to \textit{glib-2.39.2}, the return types of some functions 
(e.g., \textit{g\_hash\_table\_replace}, \textit{g\_hash\_table\_insert})
changed from \textit{void} to \textit{gboolean}.
These changes are:
1) backward compatible ---
a binary complied against the old version will ignore the return value
of the new version, and there is no error;
2) forward incompatible ---
a binary complied against the new version may use the return value,
where the old version returns void.

These changes may cause \textit{DepBugs} in many applications (e.g., \textit{cockpit-202.1}, \textit{homebank-5.2.2}), 
where the return values of the changed functions are used.
Figure~\ref{fig:example} shows code snippets of two applications.
The usage of return values indicates any \textit{glib} version returning void will be incompatible to the applications.
However, in Ubuntu-19.10, \textit{cockpit-202.1} depends on \textit{glib>=2.37.6}, and \textit{homebank-5.2.2} depends on \textit{glib>=2.37.3}.
Both the version ranges contain the incompatible version \textit{glib-2.39.1}.
Therefore, we say \textit{cockpit-202.1} and \textit{homebank-5.2.2} contain \textit{DepBugs} 
since their version ranges contain incompatible versions.

The root cause of the \textit{DepBugs} is that library developers do not update the
introducing versions of the changed functions in the \textit{symbols} file of the library.

\textbf{The \Name{} approach.}
\Name{} can detect \textit{DepBugs} in the above example,
and prevent \textit{CFailures} caused by the bugs.
Figure~\ref{fig:overview} shows the overview of \Name{}, which contains three major steps.
First, the root causes of \textit{CFailures} are incompatible changes in libraries.
\Name{} collects incompatible changes from any two successive library versions,
including both \textit{BICs} and \textit{FICs}.
For example, the above example contains two incompatible changes
as shown in Table~\ref{tab:changes}.

Second, one incompatible change may or may not result in \textit{CFailures}.
\Name{} analyzes usages of the 
changed element (e.g., \textit{g\_hash\_table\_replace}) in each application, and detects whether the
old or new library version of the change is incompatible to the application.
If yes, \Name{} reports a \textit{DepBug} when the incompatible version is included in the required version range of the application.
For the above example, the third column of Table~\ref{tab:dependencies} shows the incompatible versions 
that cause \textit{DepBugs}.
\ignore{
For the last example,
\Name{} finds the data-type is used by at least one symbol in the package, 
and the additional field is also used by the package.
Then, \Name{} determines the problem version is 3.7.6.3, which do
not support \textit{xSavepoint} as yet.
}

\begin{table}[tb]
  \caption{Examples of \Name{} results.}
  \label{tab:workflow}
  \footnotesize
 \begin{subtable}[c]{.49\textwidth}
 \centering
  \caption{Collecting incompatible changes in libraries.}
  \label{tab:changes}
  \begin{tabular}{|c|c|l|} \hline
    Library & Change Versions & Change Content \\ \hline
    glib & <2.39.1, 2.39.2> & \textit{g\_hash\_table\_replace} adds return values\\ \hline
    glib & <2.39.1, 2.39.2> & \textit{g\_hash\_table\_insert} adds return values \\ \hline
  \end{tabular}
 \end{subtable}
 \begin{subtable}[c]{.49\textwidth}
 \centering
 \vspace*{6pt}
  \caption{Detecting \textit{DepBugs} and suggesting incompatible library versions.}
  \label{tab:dependencies}
  \footnotesize
  \begin{tabular}{|c|c|c|c|} \hline
    Application & Library & \textit{DepBug} & Incompatible Versions \\ \hline
    cockpit-202.1 & glib>=2.37.6 & 2.39.1 & 2.37.6<=glib<=2.39.1 \\ \hline
    homebank-5.2.2 & glib>=2.37.3 & 2.39.1 & 2.37.3<=glib<=2.39.1 \\ \hline
 \end{tabular}
 \end{subtable}
 \vspace*{-12pt}
\end{table}

Third, one incompatible change may cause multiple incompatible versions.
\Name{} suggests all incompatible versions caused by each incompatible change.
Users can prevent \textit{CFailures} by avoiding using the incompatible versions.
In this step, any version that is both backward and forward compatible to the 
version reported by the second step (e.g., \textit{glib-2.39.1} for \textit{cockpit-202.1}) 
will also be regarded as an incompatible version.
In our example, the changed functions return void in \textit{glib-2.39.1} and previous versions.
Thus, the incompatible version range is \textit{glib<=2.39.1}.
Then, \Name{} calculates the intersection between the incompatible version range
and the required version range.
For example, the intersection for \textit{cockpit-202.1} is \textit{2.37.6<=glib<=2.39.1}.

There are three challenges in the design of \Name{}:
\begin{itemize}
\item \Name{} collects library changes that break either backward or forward compatibility,
whereas existing tools mainly focus on detecting backward incompatibilities.
To achieve this, we propose a heuristic rule to help \Name{} detect changes breaking forward compatibilities.

\item \Name{} detects if incompatible changes will cause \textit{DepBugs}.
This is challenging because the changes can involve different types (e.g., add a function,
remove a parameter). To address this, we categorize  the  changes
and derive a set of rules to detect \textit{DepBugs} for each type. 
 
\item \Name{} suggests all incompatible versions caused by each incompatible change.
This is non-trivial because multiple changes may affect the same element.
In this regard, \Name{} performs a global check across all versions 
to suggest incompatible ones for a changed element. 
\end{itemize}

%% file: 3-design.tex

\section{\Name{} Approach}\label{sec:design}

\begin{figure}[tb]
{\centering \resizebox*{1\columnwidth}{!}
{\includegraphics{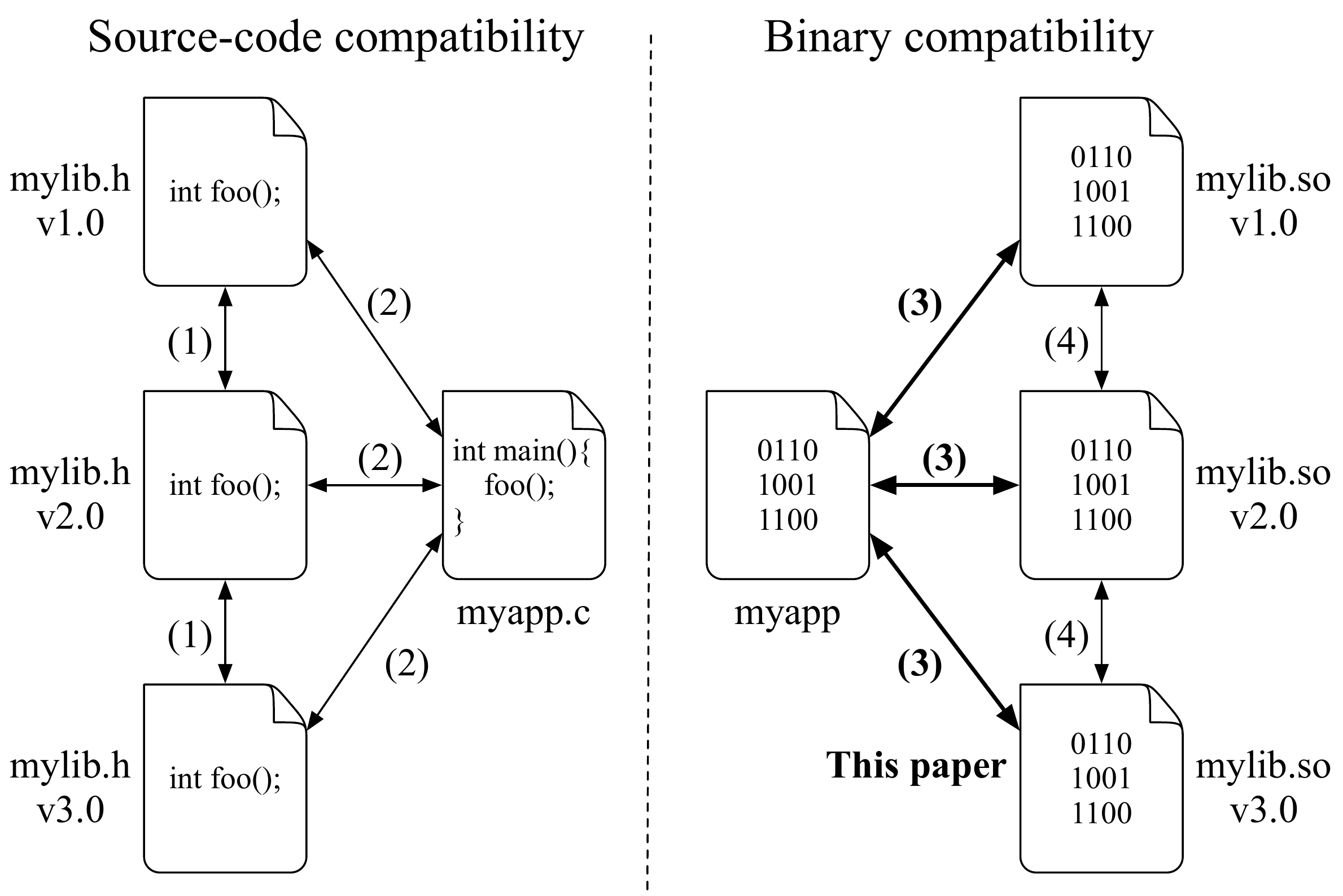}}
\caption{Difference between \Name{} and existing tools. \small{\emph{This figure includes source-code level (1, 2) and binary level (3, 4) compatibility between libraries and applications (2, 3), or cross different library versions (1, 4). Existing tools focus on (1, 2, 4), while \Name{} addresses (3).}}}
\label{fig:novelty}}
\vspace*{-12pt}
\end{figure}

There have been some existing techniques (e.g. compilers) on analyzing API usages 
in applications to check if the application is compatible with a given library version. They
work at the source-code level.  
However, end users often prefer to install binary files directly, instead of downloading source-code files and compiling the applications themselves. 
Therefore, the users often care more about the binary level compatibility.
There has also been some work (e.g. ABI Tracker~\cite{Andrey2019Abi}) on 
detecting incompatibilities cross different library versions at both source-code and binary levels.
This work does not analyze the API usages in applications. 
As shown in Figure~\ref{fig:novelty}, in this paper, we focus on detecting binary level compatibility between libraries and applications.
The compatibility at the source-code level cannot guarantee the compatibility at the binary level, such as 
modification of virtual tables of classes, change of type sizes of function parameters, change of values of enumeration elements, change of orders of struct fields, change of compilation directives, and so on.

\begin{figure}[tb]
{\centering \resizebox*{0.95\columnwidth}{!}
{\includegraphics{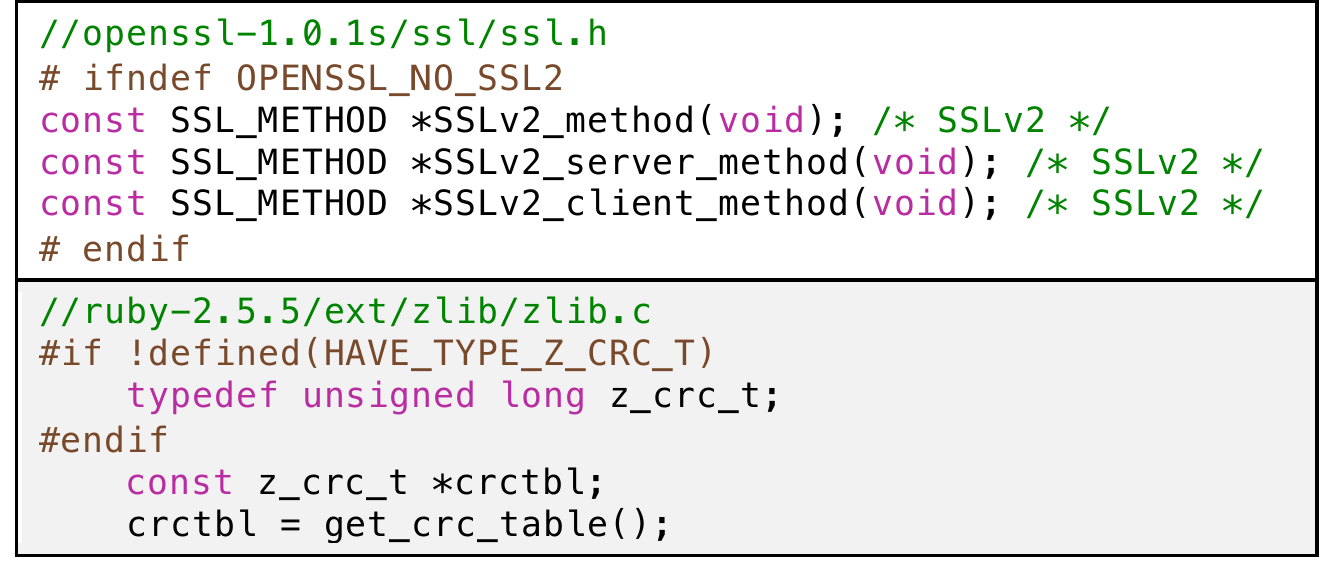}}
\caption{Examples of source-code compatible but binary incompatible dependency between libraries and applications.}
\label{fig:binary}}
\vspace*{-6pt}
\end{figure}

Figure~\ref{fig:binary} shows two real-world examples that applications and libraries are compatible at the source-code level, but incompatible at the binary level.
In the first example, three APIs in the library \textit{openssl} depend on the compilation directive OPENSSL\_NO\_SSL2.
In \textit{openssl-1.0.1s}, the directive is enabled; thus, the APIs are not available in library binaries.
While in other versions, the directive is disabled by default.
In this case, the source code of \textit{openssl-1.0.1s} is the same as the source code of other versions, 
but applications using the APIs only fail when linking to \textit{openssl-1.0.1s}.
In the second example, the application \textit{ruby-2.5.5} depends on the library \textit{zlib},
which defines \textit{z\_crc\_t} as \textit{unsigned int} after \textit{zlib-1.2.7}.
When compiling \textit{ruby} against \textit{zlib-1.2.6}, the compilation directive HAVE\_TYPE\_Z\_CRC\_T is not defined;
thus, \textit{z\_crc\_t} is \textit{unsigned long}.
When compiling \textit{ruby-2.5.5} against \textit{zlib-1.2.7}, the compilation directive is defined;
thus, \textit{z\_crc\_t} is \textit{unsigned int}.
The application \textit{ruby-2.5.5} is source-code compatible with both \textit{zlib-1.2.6} and \textit{zlib-1.2.7}.
However, when the application is compiled against one version, it will be incompatible to another version at runtime.

Algorithm~\ref{fig:pseudo} shows how \Name{} suggests incompatible versions
for each pair of library and application <$lib$, $app$> in a software repository (line 1).
\Name{} first collects the set of incompatible changes $IC$ from $lib$ (line 2).
Table~\ref{tab:changes} illustrates two examples of incompatible changes.
Each incompatible change $ic$ is a three-tuple:
<library name, change versions, change content>.
The change versions contain the old and new versions involved in the change.
For each $ic$ (line 3), \Name{} then detects whether $ic$
can cause a \textit{DepBug} in $app$,
and returns a two-tuple: <$v_{old}$, $v_{new}$> (line 4).
If the old (new) version of $ic$ is incompatible to $app$ and included in the 
version range required by $app$, $v_{old}$ ($v_{new}$) returns the old (new) version 
number, otherwise $v_{old}$ ($v_{new}$) returns -1.
If $v_{old}$ ($v_{new}$) does not return -1 (line 5, line 8), \Name{} 
will suggest any version which is both backward and forward compatible 
to $v_{old}$ ($v_{new}$) as an incompatible version (line 6, line 9).

\subsection{Collecting Incompatible Changes}\label{subsec:design1}
The first component of \Name{} takes the library $lib$ as input, 
and collects its incompatible changes $IC$.
As shown in Figure~\ref{fig:failures},
both \textit{BICs} and \textit{FICs} may result in \textit{CFailures}. 
\Name{} needs to collect both kinds of library changes.
There are existing tools of detecting compatibility problems in libraries,
e.g., \textit{ABI-Tracker}~\cite{Andrey2019Abi},
a tool for checking backward compatibility of a C/C++ library.
However, the existing tools mainly focus on backward compatibility problems.
\Name{} transfers the forward problems into backward problems.

We refer to incompatible changes from version $v_{old}$ to version 
$v_{new}$ as $IC(v_{old}, v_{new})$:
\begin{equation}
\small
\label{eqn:a}
IC(v_{old}, v_{new}) = BIC(v_{old}, v_{new}) \cup FIC(v_{old}, v_{new}),
\end{equation}
where $BIC(v_{old}, v_{new})$ and $FIC(v_{old}, v_{new})$ stand for \textit{BICs} and \textit{FICs} 
from $v_{old}$ to $v_{new}$.
\Name{} applies a heuristic rule:
forward incompatibility from $v_{old}$ to $v_{new}$ is equivalent to backward incompatibility
from $v_{new}$ to $v_{old}$, formalized as:
\begin{equation}
\small
\label{eqn:b}
FIC(v_{old}, v_{new}) = BIC(v_{new}, v_{old}).
\end{equation}
According to Equation~\ref{eqn:a} and Equation~\ref{eqn:b}, we can get:
\begin{equation}
\small
\label{eqn:c}
IC(v_{old}, v_{new}) = BIC(v_{old}, v_{new}) \cup BIC(v_{new}, v_{old}).
\end{equation}
Then, \Name{} collects both $BIC(v_{old}, v_{new})$ and $BIC(v_{new}, v_{old})$
by using the \textit{ABI-Tracker} tool.
For a library with \textit{N} versions, \Name{} calculates all incompatible changes $IC$ of $lib$:
\begin{equation}
\small
IC = \cup^{i=1}_{i=N-1}IC(v_i, v_{i+1}).
\end{equation}

During collecting library changes, 
\Name{} also consider the following factors:
1)~Library \textit{soname}~\cite{Wikipedia2019soname}.
\Name{} will skip the library changes between $v_{old}$ and $v_{new}$,
if $v_{old}$ and $v_{new}$ have different \textit{sonames}. 
Library versions with different \textit{sonames} will be packaged into different
packages; thus will not lead to \textit{DepBugs}.
2)~Symbol versioning~\cite{GNU2019VERSION}.
Symbol versioning supports multiple symbol versions in one library version.
For example, in \textit{glibc-2.27}, the symbol \textit{glob} has two versions: 
\textit{glob@@GLIBC\_2.27} and \textit{glob@GLIBC\_2.2.5} (`\textit{@@}' means the default version).
\Name{} regards symbols with different versions as different symbols.
\ignore{
3)~Change severity. 
The \textit{ABI-Tracker} tool classifies its results according to severity.
\Name{} ignores low-severity changes since they may not lead to \textit{DepBugs}.
For example, from \textit{glib-2.63.1} to \textit{glib-2.63.2}, parameter types of 
some functions change from \textit{char const*} to \textit{gchar const*}.
The change will not result in any \textit{CFailures}.
}

For each library version, \Name{} requires its binaries compiled with debug symbols.
When the input is not available, \Name{} takes source code as input, 
and compiles the library with debug symbols itself (we provide compiling scripts to achieve this).
\Name{} uses default compilation directives during the compiling process, 
and accepts custom directives provided by users at the same time.

\begin{algorithm}[tb]
\footnotesize
\caption{Pseudo-code of the \Name{} Approach.}
\label{fig:pseudo}
\begin{algorithmic}[1]
\raggedright
\Require Library set $Lib$, application set $App$
\Ensure Incompatible version sets $V_{<lib, app>}$ ($lib \in Lib, app \in App$).
\For {each pair of <$lib$, $app$>}
\State $IC$ = Collect\_Incompatible\_Change($lib$)
\For {each $ic \in IC$}
\State $[v_{old}, v_{new}]$ = Detect\_Dependency\_Bug($ic$, $app$)
\If {$v_{old} \neq -1$}
\State $V_{<lib, app>}$ += Suggest\_Incompatible\_Version($ic$, $v_{old}$, $lib$)
\EndIf
\If {$v_{new} \neq -1$}
\State $V_{<lib, app>}$ += Suggest\_Incompatible\_Version($ic$, $v_{new}$, $lib$)
\EndIf
\EndFor
\EndFor
\end{algorithmic}
\end{algorithm}

\subsection{Detecting Dependency Bugs}\label{subsec:design2}
The second component of \Name{} is to
analyze usages of the changed element of each $ic$ in $app$, and detect whether 
$v_{old}$ or $v_{new}$ is incompatible to $app$.
If yes, \Name{} reports a \textit{DepBug} when the incompatible version (i.e., $v_{old}$ or $v_{new}$)
is included in the version range required by $app$.
When $app$ does not specify any version range, \Name{} assumes it accepts all versions.
As a common usage scenario of \Name{} is to detect \textit{DepBugs} in a software repository.
In this case, \Name{} takes the repository as input, and for each application package in the repository, 
\Name{} detects whether the change can lead to a \textit{DepBug}.
It is time consuming to analyze all application packages since a software repository may 
contain tens of thousands of application packages.
In this regard, \Name{} splits the detecting process into two phases: filtering phase and detecting phase.

\textbf{Filtering phase}. \Name{} first filters out the application package that does not accept the library versions where $ic$ happened.
For example, $app$ requires \textit{libfoo>=3.0}, while the $ic$ happened from \textit{libfoo-1.0} to \textit{libfoo-2.0}.
To achieve this, \Name{} analyzes the dependencies of $app$ (e.g. from \textit{control} file in Ubuntu 
or \textit{.spec} file in Fedora),
and extracts the libraries required by $app$, as well as corresponding required version ranges.
\Name{} checks if the library (where $ic$ happens) is included in the required libraries,
and if $v_{old}$ and $v_{new}$ of $ic$ are included in the corresponding version range.
When either of the above two conditions is not satisfied, it means $ic$ can never affect $app$.
In this case, \Name{} reports no \textit{DepBugs} and stops analyzing.

Then, \Name{} filters out the application package that does not use the changed element in $ic$.
For example, the library adds a parameter for a symbol, which is not used in $app$.
In general, $ic$ can be classified into two types according to the changed 
element:
change a symbol (e.g., from "\textit{foo()}" to "\textit{foo(node a)}") and
change a data type (e.g., from "\textit{struct node \{int i;\}}" to "\textit{struct node \{float f;\}}").
\Name{} analyzes the binary files contained in $app$.
When $ic$ changes a symbol, \Name{} checks if any binary file 
requires the symbol by using the \textit{readelf}~\cite{GNU2019Readelf} tool.
When $ic$ changes a data type, \Name{} collects all symbols that use the data type 
in the library, and checks if any binary file requires any symbol.
If yes, it means $ic$ can potentially lead to \textit{CFailures},
and \Name{} starts the next phase.
Otherwise, \Name{} stops analyzing, and reports no \textit{DepBugs}.

\textbf{Detecting phase}. \Name{} analyzes the usage of the changed element and determines whether $v_{old}$ or $v_{new}$ is incompatible to $app$. 
If the change is a \textit{BIC} (\textit{FIC}) and the usage matches $v_{old}$ ($v_{new}$), 
then $v_{new}$ ($v_{old}$) will be regarded as the incompatible version.
\Name{} takes the application binary file with debug symbols as input.
When $ic$ changes a symbol, \Name{} extracts the symbol signature from the binary file.
When $ic$ changes a data type, \Name{} extracts the data-type definition from the binary file.
After that, \Name{} compares if the signature or definition is the same as that of $v_{old}$ or $v_{new}$. 
If the above input is not available, \Name{} can also extracts the usage from source code.
For example, when working on a software repository, many applications are released without debug symbols.
In this case, \Name{} automatically downloads the source code of each application package.

When using the application source code, it is hard to extract symbol signatures or data-type definitions,
since the header files are not available.
\Name{} has to apply different rules to determine the incompatible version.
For example, when $ic$ adds a field in a \textit{struct}, 
\Name{} needs to check if the additional field is used in the source code.
When $ic$ changes the type of a return value from \textit{void} to \textit{non-void}, 
\Name{} needs to check if the return value is used in the source code.

In this regard, 
we enumerate all types of incompatible changes in C/C++ libraries and define determination rules for each type.
The classification and rules are shown in Table~\ref{tab:rules}.
We classify library changes into 18 types related to \textit{enum} (1-3), 
\textit{struct} (4-7), \textit{variable} (8-10), and \textit{function} (11-18).
The \textit{struct} and \textit{enum} types are data-type changes,
while the \textit{variable} and \textit{function} types are symbol changes.
For data-type changes (1-7), \Name{} needs to confirm that the application 
uses the changed element in source code, e.g., member for \textit{enum} or
field for \textit{struct}.
For symbol changes (8-18), \Name{} has already confirmed that the application
uses the changed symbol in the filtering phase.
For changes related to "add" or "remove" (1-2, 4-5, 8-9, 11-14, 16-17), once the application 
uses the changed element, \Name{} determines the incompatible version is 
$v_{old}$ or $v_{new}$, respectively.
For changes related to "change type" (6, 10, 15, 18), \Name{} analyzes the usages 
of changed element, and infers the type in source code.
For example, from \textit{zlib-1.2.6.1} to \textit{zlib-1.2.7}, the return type of the function
\textit{get\_crc\_table} changed from \textit{long} to \textit{int}.
In the source code of package \textit{unalz-0.65}, \Name{} finds
"long *CRC\_TABLE = get\_crc\_table();",
i.e., the return type matches version 1.2.6.1. Thus,
\Name{} determines 1.2.7 is the incompatible version.
As for change type 3 and 7, it is hard to infer the member value or
field order from source code. Thus, \Name{} cannot determine the incompatible version.

We tried to build a complete table with our best effort. We referenced 
online resources during the enumeration process~\cite{Andrey2019Checker, KDE2019Policies, Josh2019ABI}. 
For example, changing an inherited class in C++ will generate two totally different symbols 
in binaries due to name mangling. 
In this case, \Name{} will report \emph{function add} and 
\emph{function remove}.
Also, \Name{} is designed to be flexible to incorporate new rules.

\Name{} uses \textit{srcML}~\cite{Michael2019srcML}, a source-code analysis infrastructure,
to achieve the above analyzing.
The source code cannot be compiled since the lack of header files, 
while srcML provides lexical analysis and syntax analysis for non-compilable source code.
\Name{} returns a two-tuple: <$v_{old}$, $v_{new}$> in this step.
If the old (new) version in $ic$ is incompatible to $app$ and included in the 
version range required by $app$, $v_{old}$ ($v_{new}$) returns the old (new) version 
number, otherwise $v_{old}$ ($v_{new}$) returns -1.

\begin{table}[tb]
\caption{Rules for determining \textit{DepBugs}.}
\label{tab:rules}
\footnotesize
\begin{tabular}{|c|l|l|c|} \hline
ID & Types of Incompatible Changes & \Name{} Rules &  \tabincell{c}{Incomp.\\Version} \\ \hline
1 & Enum adds member & Use the member & $v_{old}$ \\ \hline
2 & Enum removes member & Use the member & $v_{new}$ \\ \hline
3 & Enum changes member value & Use the member & - \\ \hline
4 & Struct adds field$^\dagger$ & Use the field & $v_{old}$ \\ \hline
5 & Struct removes field& Use the field & $v_{new}$ \\ \hline
6 & Struct changes field type & \tabincell{l}{Use the field \& \\ Match the filed type} & $v_{o.}$/$v_{n.}$ \\ \hline
7 & Struct changes field order & Use the field & - \\ \hline
8 & Global variable adds & - & $v_{old}$ \\ \hline
9 & Global variable removes & - & $v_{new}$ \\ \hline
10 & Global variable changes type & Match the var type & $v_{o.}$/$v_{n.}$ \\ \hline
11 & Function adds & - & $v_{old}$ \\ \hline
12 & Function removes & - & $v_{new}$ \\ \hline
13 & Function adds para & Use the para & $v_{old}$ \\ \hline
14 & Function removes para & Use the para & $v_{new}$ \\ \hline
15 & Function changes para type & Match the para type & $v_{o.}$/$v_{n.}$ \\ \hline
16 & Function adds return value& Use the function ret & $v_{old}$ \\ \hline
17 & Function removes return value& Use the function ret & $v_{new}$ \\ \hline
18 & Function changes return type & Match the ret type & $v_{o.}$/$v_{n.}$ \\ \hline
\end{tabular}
\begin{flushleft}
\footnotesize{
$^\dagger$ The \textit{struct} related rules (4-7) also apply for \textit{union} or \textit{class}.\\}
\end{flushleft}
\vspace*{-6pt}
\end{table}

\ignore{
\begin{table*}[tb]
\caption{Rules for determining \textit{DepBugs}.}
\vspace*{-6pt}
\label{tab:rules}
\small
\begin{tabular}{|c|l|c|c|c|c|c|} \hline
\multicolumn{2}{|c|}{}&\multirow{2}*{Add}&\multirow{2}*{Remove}&\multicolumn{3}{c|}{Change} \\ \cline{5-7}
\multicolumn{2}{|c|}{}&&&Change Type&Change Value&Change  Order\\ \hline
\multirow{4}*{Type}&Enumeration&&&&& \\ \cline{2-7}
&Enumeration Member&$v_{old}$&$v_{new}$&&& \\ \cline{2-7}
&Struct/Union/Class&&&&& \\ \cline{2-7}
&Struct/Union/Class Field&&&&& \\ \hline
\multirow{4}*{Symbol}&Global Variable&&&&& \\ \cline{2-7}
&Function&&&&& \\ \cline{2-7}
&Function Parameter&&&&& \\ \cline{2-7}
&Function Return Value&&&&& \\ \hline
\end{tabular}
\vspace*{-12pt}
\end{table*}
}

\subsection{Suggesting Incompatible Versions}
We refer to the incompatible version reported in the above step (i.e., $v_{old}$ or $v_{new}$) as $v_{bug}$.
A library change may lead to multiple incompatible versions beyond $v_{bug}$.
In this component,
\Name{} detects all library versions that are incompatible to $app$ caused by $ic$.
To achieve this,
\Name{} cannot simply assume the versions less than or greater than 
$v_{bug}$ as incompatible versions,
since the changed element in $ic$ may change again in another $ic$.
For example, in \textit{zlib}, developers remove the function \textit{gzgetc} 
(change \textit{gzgetc} to a macro for speed)
from \textit{zlib-1.2.5.1} to \textit{zlib-1.2.5.2}.
After that, the developers restore \textit{gzgetc} for compatibility
from \textit{zlib-1.2.5.2} to \textit{zlib-1.2.5.3}~\cite{Zlib2019ChangeLog}.
In this regard, \Name{} checks compatibilities of the changed element of $ic$ 
across all versions of $lib$, and any version that is both backward 
and forward compatible to $v_{bug}$ will be regarded as an
incompatible version.

We refer to the changed element in $ic$ as $ele$. 
Suppose there are \textit{N} library versions.
For $\forall i \in [1, N]$, \Name{} calculates $isIV(v_i)$, a Boolean value
indicating whether $v_i$ is an incompatible version:
\begin{equation}
\small
\label{eqn:x}
isIV(v_i) = \lnot bbc(v_{bug}, v_i, ele) \land \lnot bfc(v_{bug}, v_i, ele),
\end{equation}
where $bbc(v_{bug}, v_i, ele)$ and $bfc(v_{bug}, v_i, ele)$ return Boolean values,  
meaning if $ele$ breaks backward compatibility or breaks forward compatibility from $v_{bug}$ to $v_i$, respectively.
If yes, return 1, otherwise return 0. Similar to Section~\ref{subsec:design1}, we have:
\begin{equation}
\small
\label{eqn:y}
bfc(v_{bug}, v_i, ele) = bbc(v_i, v_{bug}, ele).
\end{equation}
Therefore, \Name{} transforms the above two equations to:
\begin{equation}
\small
isIV(v_i) = \lnot bbc(v_{bug}, v_i, ele) \land \lnot bbc(v_i, v_{bug}, ele).
\end{equation}
Then, \Name{} outputs a list of Boolean values $ISIV$, 
each of them indicates whether a version is incompatible (i.e., 1) or not (i.e., 0):
\begin{equation}
\small
ISIV = [isIV(v_1), isIV(v_2), ..., isIV(v_N)].
\end{equation}

For each element (e.g. $isIV(v_i)$) in $ISIV$, if $isIV(v_i)$ 
equals to 1, and $v_i$ belongs to the version range required by $app$, 
\Name{} regards $v_i$ as an incompatible version.
Taking the application \textit{cockpit-202.1} as an example,
the required version range is \textit{glib>=2.37.6};
while for $\forall j \in$ (\textit{glib<=2.39.1}), $isIV(v_j)$ equals to 1.
\Name{} suggests the incompatible versions are \textit{2.37.6<=glib<=2.39.1}.
For an application that is not managed in a software repository, 
\Name{} assumes that it accepts all library versions since there is no version ranges.

For the given $app$ and $lib$, \Name{} reports a set of incompatible versions
for each $ic$: $IV_{<lib,\ app,\ ic>}$. Suppose there are \textit{M} incompatible changes in $lib$.
Finally, \Name{} suggests all incompatible versions between
$app$ and $lib$:
\begin{equation}
V_{<lib,\ app>} = \cup^{i=1}_{i=M}IV_{<lib,\ app,\ ic_i>},
\end{equation}
where $ic_i$ stands for the $i_{th}$ incompatible change.

%% file: 4-evaluation.tex

\section{Evaluation}\label{sec:evaluation}


To evaluate \Name{}, we consider three research questions:

\vspace*{3pt}
\noindent
{\bf RQ1:}
How effective is \Name{} at preventing known \textit{CFailures}?
This question examines the \textit{recall} of  \Name{} by calculating the percentage of 
\textit{CFailures} that can be prevented by \Name{} among all known \textit{CFailures}.

\vspace*{3pt}
\noindent
{\bf RQ2:}
How effective is \Name{} at preventing unknown \textit{CFailures}?
This question evaluates the \textit{precision} of \Name{} by calculating the percentage of 
correct results among all results reported by \Name{}.

\vspace*{3pt}
\noindent
{\bf RQ3:} 
How does \Name{} compare with existing methods?
This question compares \Name{} with two widely used \textit{DMSs}
(i.e., \textit{apt} and \textit{dnf}), as well as the dependencies declared
in the build systems (e.g., \textit{autoconf} or \textit{cmake}) by developers\footnote{
The data and source code in this paper are publicly available in https://github.com/ZhouyangJia/DepOwl.}.
\vspace*{3pt}

\ignore{
\begin{table}[tb]
  \caption{Library packages used to detect bad dependencies.}
  \vspace*{-6pt}
  \label{tab:library}
  \small
  \begin{tabular}{|l|l|l|} \hline
libglib2.0-0        & libxml2   & libkf5coreaddons5       \\ \hline
zlib1g              & libcairo2            & libqt5network5          \\ \hline
libqt5core5a        & libssl1.1            & libpango-1.0-0                \\ \hline
libgmp10            & libmpfr6       & libkf5configcore5       \\ \hline
libx11-6            & libjpeg8              & libqt5widgets5             \\ \hline
libqt5gui5          & libtinfo6            & libgdk-pixbuf2.0-0                \\ \hline
libqt5dbus5      & libpng16-16          & dconf-gsettings-backend \\ \hline
libgtk-3-0          & libkf5i18n5          & libpangocairo-1.0-0                \\ \hline
libxext6            & libsqlite3-0         & libsdl1.2debian         \\ \hline
libisl21            & libmpc3  &  libkf5widgetsaddons5             \\ \hline
libasound2 & libqt5xml5           &                       \\  \hline
\end{tabular}
\vspace*{-12pt}
\end{table}
}

\begin{table*}[tb]
\footnotesize
\centering
\caption{Examples of reported \textit{DepBugs} in the software repository shipped with Ubuntu-19.10 $^\dagger$.}
\label{tab:sample}
\begin{tabular}{|c|c|c|l|c|} \hline  
  \multicolumn{2}{|c|}{Application and Library Information} & \multicolumn{3}{|c|}{Results of \Name{}} \\ \hline
  Application Package & Library Package & Change Versions & Change Symbol/Data-type& Incompatible Versions \\ \hline
  qgis-providers\_3.4.10 & libsqlite3-0>=3.5.9 & <3.7.6.3, 3.7.7> & \textit{struct sqlite3\_module} adds \textit{xSavepoint} & [3.5.9, 3.7.6.3] \\ \hline
  unalz\_0.65-7 & zlib1g>=1.1.4 & <1.2.6.1, 1.2.7> & \textit{get\_crc\_table} changes return value from long to int& [1.2.7, $V_{last}$] \\ \hline
  elisa\_1.1 & libkf5i18n5>= 5.15.0 & <5.16.0, 5.17.0> & Add \textit{KLocalizedContext(QObject*)} & [5.16.0] \\ \hline
  gammaray\_2.9.0 & libqt5core5a>=5.12.2 & <5.13.2, 5.14.0> & \textit{qt\_register\_signal\_spy\_callbacks()} changes para type & [5.14.0, $V_{last}$] \\ \hline
  geeqie\_1:1.5.1-1 & libglib2.0-0>=2.51.0 & <2.51.0, 2.52.0> & \textit{g\_utf8\_make\_valid()} adds parameter \textit{gssize} & [$V_{init}$, 2.51.0] \\ \hline
  alsa-utils\_1.1.9 & libasound2>=1.1.1 & <1.1.9, 1.2.1> & Remove \textit{snd\_tplg\_new@ALSA\_0.9} & [1.2.1, $V_{last}$] \\ \hline
  rkward\_0.7.0b-1.1 & libkf5coreaddons5>=5.19.0 & <5.19.0, 5.20.0> & Add \textit{KCoreAddons::versionString()}  & [5.19.0] \\ \hline
\end{tabular}
\begin{flushleft}
\footnotesize{$^\dagger$ We illustrate one bug for each library package.
The complete \textit{DepBug} list is available in our supplementary materials. }
\end{flushleft}
\vspace*{-6pt}
\end{table*}

\subsection{Datasets and Experiment Designs}
\label{subsec:dataset}
For each research question, we introduce the preparation of datasets,
and the measurements used during the evaluation.

\textbf{RQ1: Preventing known \textit{CFailures}.}
We collected known \textit{CFailures} from StackOverflow by using
keyword search.
However, simple keywords (e.g.,  library, dependency, version, etc) may 
result in tens of thousands of issues, and introduce massive manual 
efforts in the following analysis.
Instead, we used the error messages when users came across 
compatibility problems as keywords.
For example, when a library removes a symbol, the 
application will echo "symbol lookup error" at runtime.
When a library symbol adds or removes a parameter, the 
complier will complain "too few/many parameter to function" at compiling time.
In total, we collected 529 issues by using error-message searching.

We then manually analyzed root causes of these issues and found
69 issues involve incompatible changes in libraries.
These changes lead to \textit{CFailures} through misuses of library versions.
While others are mainly caused by dependency problems but not related to compatibility.
Among the 69 issues, 
38 of them involve C/C++ programs. Since the current version of \Name{}  handles C/C++ 
programs,  we used the 38 issues to answer RQ1. 
The applications of 23 issues are code snippets provided by the original posters, 
while other issues involved 12 mature projects including servers (e.g., Httpd, MongoDB) 
and clients (e.g., Eclipse, Qt) from different domains. 

Since the 38 issues were selected by searching error messages, they may 
not cover certain types of compatibility breaking 
changes (Table~\ref{tab:rules}) that do not produce observable symptoms. 
For example, in Table~\ref{tab:rules},  ``changing member values in a enum type (ID 3)"
and ``changing field orders in a struct type (ID 7)" may result in errors in a program, 
but will not generate error messages.
Therefore, the 38 issues cannot cover the changes of ID 3 and ID 7.
It is hard to collect incompatibilities that have no observable failures,
since users cannot be sure if they are actual bugs, thus may not report issues.

We measured the effectiveness of preventing known \textit{CFailures} in terms of whether
\Name{} can prevent the \textit{CFailures} in the 38  C/C++ related issues.
To achieve this, \Name{} needs to detect \textit{DepBugs} in these issues.
\textit{DepBugs} happen when the version ranges required by applications contain incompatible versions.
Fixing the \textit{DepBugs} helps users avoid using incompatible versions and prevent \textit{CFailures}.
When an application does not specify a version range,
\Name{} assumes that the application accepts all library versions.

\textbf{RQ2: Preventing unknown \textit{CFailures}.}
We used the software repository shipped with Ubuntu-19.10 (the latest stable version at the time of writing)
to evaluate \Name{},
since Ubuntu uses \textit{apt}, which can resolve dependencies automatically, 
while other \textit{DMSs} mainly depend on application developers to manually input dependencies.
The repository includes 61,068 packages;
each package can be either an application package or a library package.
There are 32,069 library packages, which are depended by at least one other package. 
For each library package, we count the number of application 
packages that depend on it.
We choose the top 1\permil\ (i.e., 32) library packages,
which are from 26 different libraries (one library may generate 
multiple packages, e.g., the \textit{qt} library generates \textit{libqt5core5a}, 
\textit{libqt5gui5} etc.).
For each chosen library, we collect its versions released during about last ten years,
and get 841 versions in total (i.e., 32.2 versions for each library on average).

It is hard to directly measure the effectiveness of preventing unknown \textit{CFailures},
since the unknown \textit{CFailures} do not happen as yet.
Instead, we measure the effectiveness in terms of whether 
\Name{} can detect unknown \textit{DepBugs} in the software repository,
and prevent potential \textit{CFailures} caused by the \textit{DepBugs}.
In specific, for each application package from the software repository, \Name{} detects whether there are
\textit{DepBugs} with regard to the chosen library packages, i.e., the version ranges required 
by the application package contain incompatible versions.
If yes, \Name{} suggests the incompatible versions that may cause \textit{CFailures}.

\textbf{RQ3: Comparing with existing methods.}
We used the same dataset in RQ1 to compare \Name{} with existing methods,
and calculated the percentage of issues that can be prevented if 
the original posters use existing methods. 

We first compared \Name{} with two \textit{DMSs} used in industry: 
1) \textit{dnf}, used in RPM-based Linux distributions,
where application developers manually specify version ranges of required libraries;
2)  \textit{apt}, used in DEB-based Linux distributions,
where library developers maintain a \textit{symbols} file.
We then compared \Name{} with building scripts (e.g., configure.ac or CMakeList.txt) shipped with application source code,
since developers often declare version ranges in the scripts.

\ignore{The techniques that are closely related to \Name{} are the \textit{DMSs} used in industry.
In additional, the building scripts (e.g., configure.ac or CMakeList.txt) shipped with application source code
can also filter out incompatible library versions.
We used two \textit{DMSs} and the build scripts as baselines to assess if \Name{} is more effective in preventing
\textit{CFailures}.
The first baseline is \textit{dnf/yum} used in RPM-based Linux distributions,
where application developers manually specify the version range of each required library.
The second baseline is \textit{apt/apt-get} used in DEB-based Linux distributions,
where library developers maintain a \textit{symbols} file.
For each application package, the version range of each required library is derived automatically.
The third baseline is the version ranges declared in the building scripts shipped with application source code.
}

\subsection{Results and Analysis}\label{subsec:result}

\textbf{RQ1: Preventing known \textit{CFailures}.}
Two authors manually evaluated whether \Name{} can prevent the 38 known \textit{CFailures}
by analyzing if the incompatible versions suggested by \Name{} 
contain the incompatible version used by the original poster.
The result shows \Name{} successfully suggests incompatible versions for 35 of the 38 C/C++ related issues.
The complete list of these issues is available in our supplementary materials.
Each issue in the list contains the issue ID, the application name, the library name, and the incompatible versions suggested by \Name{}.
Taking issue 27561492 as an example, library \textit{libpcre} adds 
function \textit{pcrecpp::RE::Init} from \textit{libpcre-5.0} to \textit{libpcre-6.0}, and
changes its parameter type from \textit{libpcre-6.7} to \textit{libpcre-7.0}.
Therefore, \Name{} reports two library changes.
Meanwhile, the application \textit{mongodb-2.4} uses \textit{pcrecpp::RE::Init},
and the parameter type is the same as the type from \textit{libpcre-6.0} to 
\textit{libpcre-6.7}.
Thus, \Name{} reports [$V_{init}$, 5.0]$\cup$[7.0, $V_{last}$]\footnote{$V_{init}$ and $V_{last}$ stand for the first and the last library version that have the same \textit{soname}.}
as the incompatible versions.

On the other hand, \Name{} reported three false negatives.
Two cases were caused by compilation directives, 
e.g., the original poster executed and compiled an application on different OS, 
where the libraries may be compiled with different directives. 
\Name{} cannot infer such directives, and thus generates false negatives. 
The last case missed version information and might have used a very old library version. 
\Name{} can prevent \textit{CFailures} in 35 out of the 38 issues.
This result indicates \Name{} can effectively prevent real-world \textit{CFailures} in terms of recall.

\textbf{RQ2: Preventing unknown \textit{CFailures}.}
\Name{} collected 27,413  incompatible changes from the 841 versions of the 26 libraries.
For each change, \Name{} detects if the change can cause a \textit{DepBug} for each application package.
\Name{} detected 77 \textit{DepBugs},
of which 49 are caused by backward incompatible changes
and 28 are caused by forward incompatible changes.
These \textit{DepBugs} involve 69 application packages and 7 library packages.
%
Table~\ref{tab:sample} illustrates one bug for each library package.
The complete \textit{DepBug} list is available in our supplementary materials. 
For example, in the first bug, the application \textit{qgis-providers\_3.4.10} 
depends on the library \textit{libsqlite3-0>=3.5.9}, which adds the filed
\textit{xSavepoint} in \textit{struct sqlite3\_module} from 3.7.6.3 to 3.7.7.
The application used the new filed; thus 3.7.6.3 is an incompatible version.
\Name{} then suggests all incompatible versions: [3.5.9, 3.7.6.3].

We searched evidence from new library versions, new application versions, or software repositories
to evaluate if the 77 \textit{DepBugs} have been handled in different ways.
If not, we further reported them to the repository maintainers.
Among the 77 \textit{DepBugs}, library developers undo the library changes of 37 cases in later library version.
It means applications may have \textit{CFailures} when using the library versions before undoing the changes.
Application developers update the application to adapt the changes in 3 cases, meaning the old application version may have \textit{CFailures}.
Besides, 24 \textit{DepBugs} are fixed in the latest version of Ubuntu or Debian.
Although these bugs have been handled in different ways, 
they had been in the system for a long period of time, posing threats to the system reliability. 
For example, library developers fixed an incompatible version, 
which had already been released and affected applications. 
\Name{} is able to prevent these impacts from the very beginning.

For the other 13 cases, we report them to the Ubuntu community, 4 of them have been confirmed by developers, and 8 are pending for response.
So far, we only found one potential false-positive case. 
\Name{} reported that the library \textit{kcoreaddons-5.19} is incompatible to the application \textit{rkward}, 
which depends on \textit{kcoreaddons>=5.19}. 
The developer agreed that the incompatibility may exist, 
but \textit{kcoreaddons-5.19} is not actually used in any Ubuntu release 
(Xenial uses kcoreaddons-5.18, Bionic uses kcoreaddons-5.40), thus has zero impact.
This result indicates \Name{} can effectively detect real-world \textit{DepBugs} in terms of precision.

This experiment took about 30 hours in a virtual machine with a dual-core CPU and 4G memory. 
The filtering and detection phases took about five hours (excluding downloading packages).
The majority of  time was spent on collecting library changes of history versions.
This process is one-time effort, since the latest library version can be analyzed incrementally.
The execution time of each library depends on its scale and type. 
When analyzing large C++ libraries like Qt, \Name{} may need dozens of minutes for each pair of versions. Meanwhile, some other libraries only need several seconds.

\begin{figure}[tb]
{\centering \resizebox*{0.75\columnwidth}{!}
{\includegraphics{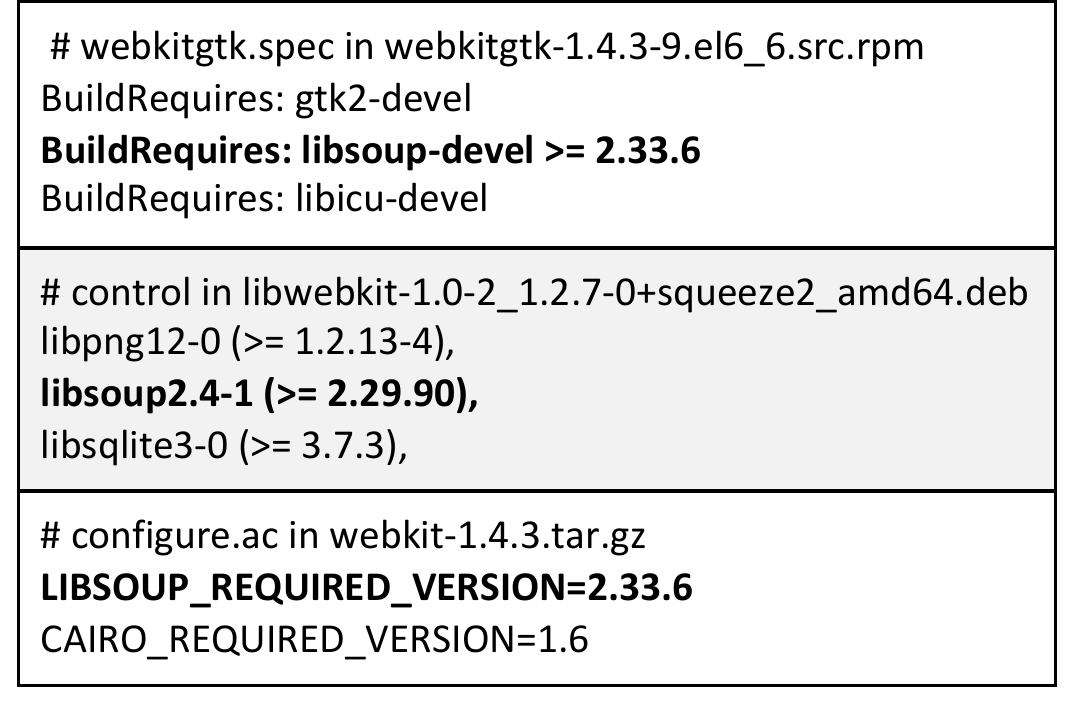}}
\caption{Version ranges of different baselines.}
\label{fig:range}}
\vspace*{-6pt}
\end{figure}

\textbf{RQ3: Comparing with existing methods.}
We compared \Name{} with three existing methods, i.e., \textit{dnf} for \textit{.rpm} packages, \textit{apt} for \textit{.deb} packages, and the building system.
For each StackOverflow issue used in RQ1, two authors manually evaluated if the \textit{CFailure} can be
prevented by using existing methods when the original poster used the existing methods at first.
Taking issue 30594269 as an example, \textit{webkit} has "symbol lookup error" when linking to \textit{libsoup}.
The incompatible version range of \textit{libsoup} is [$V_{init}$, 2.29.6].
The version ranges of \textit{libsoup} in three baselines accepted by \textit{webkit}
are [2.33.6, $V_{last}$], [2.29.90, $V_{last}$], [2.33.6, $V_{last}$], respectively.
Thus, all the three baselines can prevent the failure in this issue.
Figure~\ref{fig:range} lists the files where we get these version ranges, including
the \textit{webkitgtk.spec} file in the \textit{.rpm} package, 
the \textit{control} file in the \textit{.deb} package, and the \textit{configure.ac} file in the building system of source code. 

\begin{figure}[tb]
{\centering \resizebox*{0.9\columnwidth}{!}
{\includegraphics{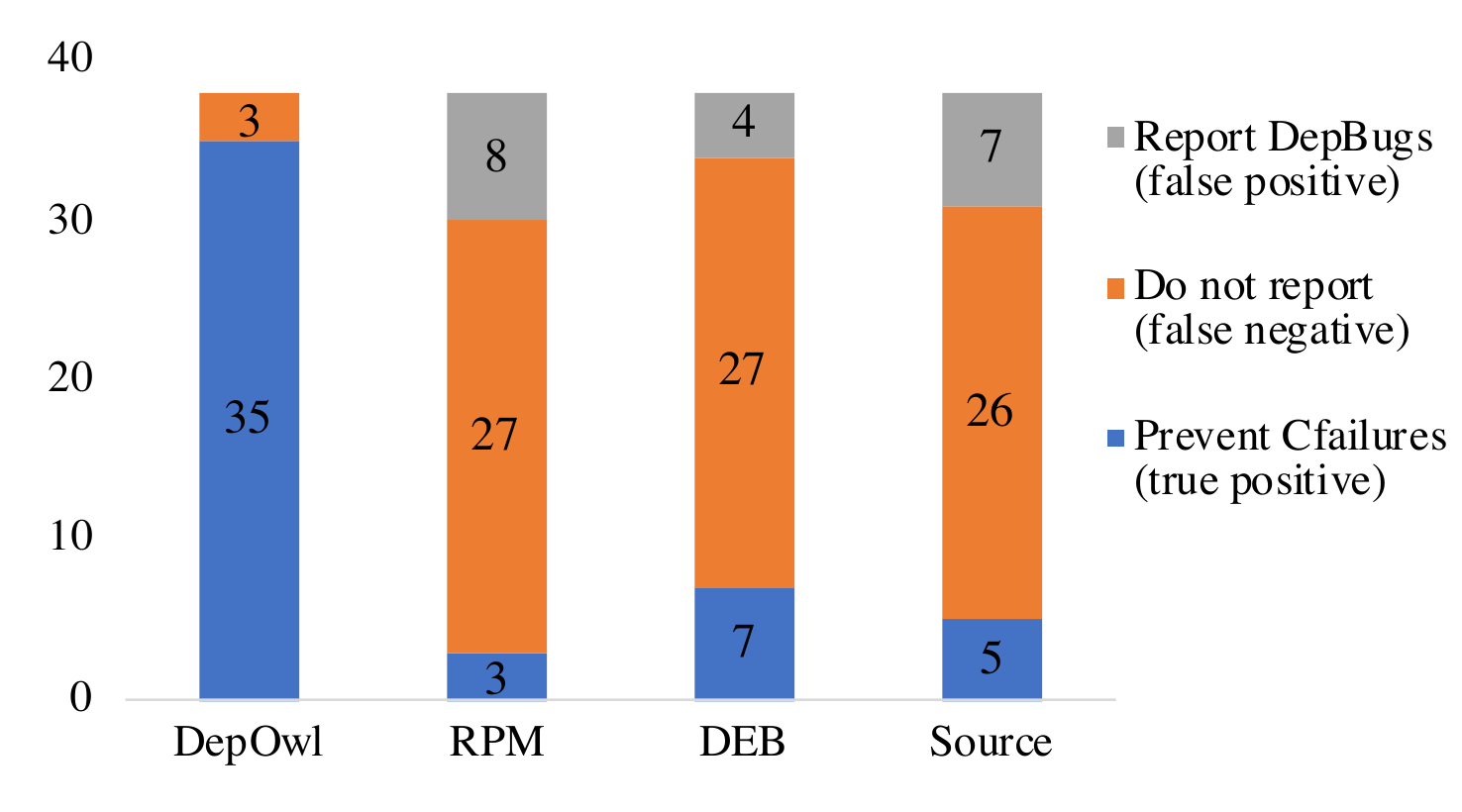}}
\caption{The comparison among \Name{} and baselines.}
\label{fig:compare}}
\vspace*{-9pt}
\end{figure}

Figure~\ref{fig:compare} shows the results regarding the comparison
among \Name{} and the three baselines.  
\Name{} can prevent \textit{CFailures} in 35 issues whereas the baselines can prevent \textit{CFailures} in 3, 7, 5 issues, respectively.
Besides, \Name{} does not report any problems in 3 issues (i.e., 3 false negatives), 
while the baselines do not report any problems in 27, 27, 26 issues.
This is because 23 issues were caused by code snippets provided by the original posters.
These code snippets are not managed in any \textit{DMSs} or build systems.
Last but not the least, the baselines report \textit{DepBugs} in 8, 4, 7 issues (i.e., the version range contains incompatible versions).
\Name{} successfully prevents 35 \textit{CFailures}, whereas the best baseline prevents 7 \textit{CFailures}.
The detailed results are available in our supplementary materials.
This result indicates \Name{} is more accurate than the three baselines. 
In the mean time, \Name{} requires no human efforts, while the baselines require
manual inputs from either library developers or application developers.

%% file: 5-discussion.tex

\section{Discussion and Future Works}\label{sec:discussion}
In this section, we discuss limitations in the design of \Name{}, 
as well as future works with regard to the limitations.

\textbf{Debug symbols of libraries.}
To collect incompatible changes (in Section~\ref{subsec:design1}), \Name{} requires all versions
of the library as inputs.
Each version should be in the source code form or the binary form with debug symbols.
For the binary form, most libraries are released without debug symbols, and do not meet 
the requirement of \Name{}.
As for the source code form, we need to compile the source code so that \Name{} can 
collect Application Binary Interface (ABI) changes.
\Name{} provides scripts to automate the compiling process.
This is still limited since \Name{} uses the default compilation directives;
thus cannot collect ABI changes triggered by other directives.
As a result, developers have to provide the compilation directives, or \Name{} may cause false negatives.

\textbullet\ \textit{Future work:}
The most convenient way to avoid this limitation is to suggest library developers to release 
binaries with debug symbols when releasing new versions.
This practice actually has been applied in some libraries.
For example, in the software repository of Ubuntu-19.10,
there are 753 packages with the suffix `-dbg' containing debug symbols.

\textbf{Code analysis in applications.}
When detecting dependency bugs (in Section~\ref{subsec:design2}), 
\Name{} requires application binaries compiled with debug symbols.
This input is not available in most applications managed in existing \textit{DMSs}.
Alternatively, \Name{} has to use source code as input, but correct usages in source code 
do not indicate the application is free of \textit{CFailures} in the binary form.
For example, the second example of Figure~\ref{fig:binary} shows the usage of \textit{get\_crc\_table} in \textit{ruby-2.5.5},
which works well against both \textit{zlib-1.2.6} and \textit{zlib-1.2.7} in source code level:
when \textit{ruby-2.5.5} is compiled against \textit{zlib-1.2.7}, the return type \textit{z\_crc\_t}
is \textit{int};
when \textit{ruby-2.5.5} is compiled against \textit{zlib-1.2.6}, the return type  \textit{z\_crc\_t}
is \textit{long}.
However, \textit{ruby-2.5.5} may have \textit{CFailures} when compiled against one version
and linked to another version at runtime.
This limitation will lead to false negatives.

\textbullet\ \textit{Future work:}
\Name{} will provide an interface for application developers to indicate a 
fixed version for each library. This manual effort is the same to most
\textit{DMSs} like \textit{pip} or \textit{Maven}.
Thus, \Name{} can compile the source code against the fixed library version.

\textbf{Limitations when using ABI-Tracker.}
\Name{} uses ABI-Tracker to collect incompatible changes of a target library.
ABI-Tracker takes source code of the library history versions as inputs
and compiles each version with default directives.
This process may introduce both false positives and false negatives.
For example, in the first example of Figure~\ref{fig:binary}, 
ABI-Tracker reports that \textit{openssl-1.0.1s} removes three symbols.
However, users will not encounter failures when disabling OPENSSL\_NO\_SSL2.
In this case, \Name{} may report false positives, although no false positives
directly related to ABI-Tracker are generated in our experiment.
On the other hand, when incompatible changes can only be triggered by specific directives,
ABI-Tracker may generate false negatives and thus cause \Name{} to report false negatives.
For example,  two out of three false negatives in RQ1
are caused by compilation directives not correctly identified by 
ABI-Tracker. 

\textbf{Impacts of compilation directives.}
The compilation directives of a target library may affect the symbols and data types provided by the library,
and further affect the results of \Name{}.
Since ABI-Tracker uses default compilation directives to compile
each library version, it may cause \Name{} to report
false negatives (as discussed in the above paragraph).
We have mitigated this impact by directly analyzing the
binaries of the target libraries without the need of providing compilation directives. 
In the case where binaries are not available, 
\Name{} accepts the directives from users for compiling. 
In our evaluation, we manually input the directives in most cases. 
For the two cases that we cannot obtain the directives in RQ1, 
\Name{} reports two false negatives, since the directives are hard to be inferred automatically.

%% file: 6-relatedwork.tex
\section{Related Works}\label{sec:work}
We briefly classify the existing works into three types:

\textbf{Library changes.}
Many works are targeted at library changes.
Bagherzadeh~\textit{et al.}~\cite{Bagherzadeh2018} studied the size, type and bug fixes 
in 8,770 changes that were made to Linux system calls.
Brito~\textit{et al.}~\cite{8330214} identified 59 breaking changes and asked the developers 
to explain the reasons behind their decision to change the APIs.
%
%
Dig~\textit{et al.}~\cite{doi:10.1002/smr.328, 1510134} discovered that over 80\% of changes that break 
existing applications are refactorings.
%
%
Li~\textit{et al.}~\cite{7816486} investigated the Android framework source code, 
and found inaccessible APIs are common and neither forward nor backward compatible.
Li~\textit{et al.}~\cite{10.1145/3196398.3196419} and Wang~\textit{et al.}~\cite{Wang2020Exploring} 
studied API deprecation in the Android ecosystem and Python libraries.
McDonnell~\textit{et al.}~\cite{6676878} found Android updates 115 API per month,
and 28\% usages in client applications are outdated with a median lagging of 16 months.
Sawant~\textit{et al.}~\cite{8453124} investigated why API producers deprecate features, 
whether they remove deprecated features,  and how they expect consumers to react.
Brito~\textit{et al.}~\cite{8330249} identified API breaking and non-breaking changes between 
two versions of a Java library.
Foo~\textit{et al.}~\cite{10.1145/3236024.3275535} presented a static analysis to 
check if a library upgrade introduces an API incompatibility.
%
%
Meng~\textit{et al.}~\cite{10.5555/2337223.2337265} aggregated the revision-level rules to obtain 
framework-evolution rules.
Mezzetti~\textit{et al.}~\cite{mezzetti_et_al:LIPIcs:2018:9212} proposed type regression testing to determine 
whether a library update affects its public interfaces.
Ponomarenko~\textit{et al.}~\cite{Ponomarenko2012} presented a new method for automatic detection 
of backward compatibility problems at the binary level.
%
%
Wu~\textit{et al.}~\cite{6062100} proposed a hybrid approach to identify framework evolution rules.

These works are targeted at detecting changes, refactorings and rules during library evolutions.
While \Name{} is targeted at preventing failures caused by the results of these works.

\textbf{Application failures.}
Some works focus on \textit{CFailures} in applications.
Cai~\textit{et al.}~\cite{10.1145/3293882.3330564} studied compatibility issues 
in 62,894 Android app to understand the symptoms and causes of these issues.
Cossette~\textit{et al.}~\cite{10.1145/2393596.2393661}  studied techniques 
to help migrate client code between library versions with incompatible APIs.
Dietrich~\textit{et al.}~\cite{6747226} studied partially upgrading systems,
and found some crucial verification steps are skipped in this process.
%
%
Jezek~\textit{et al.}~\cite{JEZEK2015129} studied the compatibility of API changes, 
and the impact on programs using these libraries.
Lamothe~\textit{et al.}~\cite{10.1145/3196398.3196420} reported their experience migrating the 
use of Android APIs based on documentation and historical code changes.
Linares-V\'{a}squez~\textit{et al.}~\cite{10.1145/2491411.2491428} studied how the fault- and 
change-proneness of APIs relates to applications' lack of success.
Xavier~\textit{et al.}~\cite{7884616} conducted a large-scale study on historical 
and impact analysis of API breaking changes.
Balaban~\textit{et al.}~\cite{10.1145/1094811.1094832} presented an approach to support client refactoring 
for class library migration.
He~\textit{et al.}~\cite{10.1145/3238147.3238185} and Xia~\textit{et al.}~\cite{Xia2020How} 
studied API compatibility in Android.
Henkel~\textit{et al.}~\cite{1553570} captured API refactoring actions, and
users of the API can then replay the refactorings to bring their client software components up to date.
Jezek~\textit{et al.}~\cite{6619503}  proposed an approach that analyses the byte-code of Java classes 
to find type inconsistencies cross components.
Li~\textit{et al.}~\cite{10.1145/3213846.3213857} proposed a approach for modeling the lifecycle of the Android APIs,
and analyzing app that can lead to potential compatibility issues.
Perkins~\textit{et al.}~\cite{10.1145/1108792.1108818} proposed a technique to generate client refactorings,
by replacing calls to deprecated methods by their bodies.
Wang~\textit{et al.}~\cite{8812128} proposed an automated approach that generates tests and collects 
crashing stack traces for Java projects subject to risk of dependency conflicts.
Xing~\textit{et al.}~\cite{4359473} recognized the API changes of the reused framework,
and proposed plausible replacements to the obsolete API based on working examples.

These works focus on detecting incompatible API usages and helping applications 
co-evolve with library evolutions, so that the latest application version works well.
While \Name{} can prevent \textit{CFailures} for users' in-use versions.

\textbf{Application-library dependencies.}
There are many works address application-library dependencies.
Bavota~\textit{et al.}~\cite{Bavota2015} studied the evolution of dependencies between projects 
in the Java subset of the Apache ecosystem.
Bogart~\textit{et al.}~\cite{10.1145/2950290.2950325} studied three software ecosystems to 
understand how developers make decisions about change and change-related costs.
Decan~\textit{et al.}~\cite{8721084} compared semantic-versioning compliance of four software packaging ecosystems, 
and studied how this compliance evolves over time.
Decan~\textit{et al.}~\cite{Decan2019} analyzed the similarities and differences between the evolution 
of package dependency networks.
Derr~\textit{et al.}~\cite{10.1145/3133956.3134059} studied library updatability in 1,264,118 apps, 
and found 85.6\% libraries could be upgraded by at least one version.
Dietrich~\textit{et al.}~\cite{10.1109/MSR.2019.00061} studied developers' choices between fixed version 
and version range from 17 package managers. 
%
%
Jezek~\textit{et al.}~\cite{6928807} provided evidences that four types of problems 
caused by resolving transitive dependencies do occur in practice.
Kikas~\textit{et al.}~\cite{10.1109/MSR.2017.55} analyzed the dependency network structure and 
evolution of the JavaScript, Ruby, and Rust ecosystems.
Kula~\textit{et al.}~\cite{Kula2018} studied 4,600 GitHub projects and 2,700 library dependencies
to understand if developers update their library.
Mirhosseini~\textit{et al.}~\cite{10.5555/3155562.3155577} studied 7,470 GitHub projects to
understand if automated pull requests help to upgrade out-of-date dependencies.
Pashchenko~\textit{et al.}~\cite{10.1145/3239235.3268920} studied whether dependencies of 
200 OSS Java libraries are affected by vulnerabilities.
Raemaekers~\textit{et al.}~\cite{6975655} investigated semantic versioning,
and found one third of all releases introduce at least one breaking change.
Xian~\textit{et al.}~\cite{Xian2020Automated} conducted an experience paper to evaluate 
existing third-party library detection tools.
Wang~\textit{et al.}~\cite{10.1145/3236024.3236056} conducted an empirical study on dependency 
conflict issues to study their manifestation and fixing patterns.
Zerouali~\textit{et al.}~\cite{10.1007/978-3-319-90421-4_6} analyzed the package update 
practices and technical lag for the npm distributions.

These works mainly assist people in understanding application-library dependencies.
While \Name{} is the first research work to 
help users avoid incompatible application-library dependency automatically.
Huang~\textit{et al.}~\cite{Huang2020Interactive} and Wang~\textit{et al.}~\cite{Wang2020Watchman} 
designed tools to detect dependency conflicts for Maven and PyPI ecosystems.
These tools focused on the diamond dependency problem,
which detects conflicts among different dependencies. 
They assume each dependency itself is correct, whereas \Name{} detects bugs within dependencies.

%% file: 7-conclusion.tex

\section{Conclusion}\label{sec:conclusion}
In this paper, we find \textit{CFailures} 
are caused by using incompatible library versions, which are hard to be prevented by the 
existing research works or industrial \textit{DMSs}.
To fill this gap, we design and implement \Name{}, a practical tool to 
prevent \textit{CFailures} by avoiding incompatible versions.
\Name{} can detect unknown \textit{DepBugs} in the software repository shipped with Ubuntu-19.10,
and prevent \textit{CFailures} in real-world issues collected from StackOverflow.
However, \Name{} still has limitations in practice. With limited helps from 
library developers (release binaries with debug symbols) 
and application developers (provide one required library version), 
\Name{} could achieve higher accuracy.
As a result, applications could be both flexible for library evolutions 
and reliable for \textit{CFailures}.